\documentclass[manuscript=article]{achemso}

\usepackage[version=3]{mhchem} % Formula subscripts using \ce{}
\usepackage{comment}
\usepackage{amssymb}
\usepackage{fourier}
\usepackage{nicefrac}
\usepackage{sidecap}
\usepackage{setspace} %%besserer marginpar: \mnote

\usepackage{lineno}
\usepackage{bm}
\author{Peter M. Piechulla}
\affiliation[MLU]
{Institute of Physics, Martin Luther University Halle-Wittenberg, 06120 Halle, Germany}
\author{Bodo Fuhrmann}
\affiliation[CMAT]
{Interdisciplinary Center of Materials Science, Martin Luther University Halle-Wittenberg, 06120 Halle, Germany}
\author{Evgeniia Slivina}
\affiliation[KIT]
{Institute of Theoretical Solid State Physics, Karlsruhe Institute of Technology, 76131 Karlsruhe, Germany}
\author{Carsten Rockstuhl}
\affiliation[KIT]
{Institute of Theoretical Solid State Physics, Karlsruhe Institute of Technology, 76131 Karlsruhe, Germany}
\alsoaffiliation[KITINT]
{Institute of Nanotechnology, Karlsruhe Institute of Technology, 76021 Karlsruhe, Germany}
\author{Ralf B. Wehrspohn}
\affiliation[MLU]
{Institute of Physics, Martin Luther University Halle-Wittenberg, 06120 Halle, Germany}
\alsoaffiliation[FhG]
{Fraunhofer Society, 80686 Munich, Germany}
\author{Alexander N. Sprafke}
\affiliation[MLU]
{Institute of Physics, Martin Luther University Halle-Wittenberg, 06120 Halle, Germany}
\email{alexander.sprafke@physik.uni-halle.de}
\title{Tailored Light Scattering through Hyperuniform Disorder in self-organized arrays of high-index Nanodisks}
\abbreviations{IR,NMR,UV}

\begin{document}



\small{keywords: \textit{tailored disorder, hyperuniformity, nanophotonics, metasurface.}}

\begin{abstract}
Arrays of nanoparticles exploited in light scattering applications commonly only feature either a periodic or a rather random arrangement of its constituents.
For the periodic case, light scattering is mostly governed by the strong spatial correlations of the arrangement, expressed by the structure factor. 
For the random case, structural correlations cancel each other out and light scattering is mostly governed by the scattering properties of the individual scatterer, expressed by the form factor. 
In contrast to these extreme cases, we show here, for the first time, that hyperuniform disorder in self-organized large-area arrays of high refractive index nanodisks enables both structure and form factor to impact the resulting scattering pattern, offering novel means to tailor light scattering. The scattering response from our nearly hyperuniform interfaces can be exploited in a large variety of applications and constitutes a novel class of advanced optical materials.  
\end{abstract}

\section{Introduction}
Micro- and nanostructured interfaces for electromagnetic scattering continue to play a major role in photonics due to their particularly wide range of existing as well as prospective applications, e.g. for light management in solar cells\cite{spinelli2012,Otto2015,Schneider2020} and solid-state lighting devices\cite{Son2012,Park2020}, spectral molecular fingerprint detection\cite{Ding2016,Langer2020}, or colored surfaces in architecture and design\cite{Chung2012,Blaesi2017,Liu2020}.
The key challenge in ongoing research is the ability to tailor the optical response from these interfaces in a quite sophisticated manner to meet the demands of specific applications. 
This fundamental scientific goal is accompanied by the engineering challenge of scalable fabrication of nanophotonic interfaces.
As today’s manufacturing capabilities fall far behind proposed theoretical approaches due to the costs and complexity of their fabrication, prevalent structures are those of highest feasibility instead, e.g. in crystalline silicon solar cells pyramidal surface textures by alkaline etching still are industry standard.\cite{Wilson_2020}
While advanced techniques such as electron beam lithography provide high flexibility in terms of producing nanostructures with quite a deterministic geometry, these fabrication technologies are generally slow and expensive, thus, e.g. applied to solar cells at most feasible for proof-of-concept or research purposes.

In this work, we focus on the implementation of photonic scattering interfaces consisting of an array of identical scatterers. 
The angle-resolved scattering ($ARS$) response of such a structure generally depends (a) on the arrangement of the scatterers, e.g. regularity of the array and distances between the scatterers, as well as (b) on the properties of the scatterers it is made of, e.g. size, shape, and involved materials.
The first property is described by the structure factor $S$, the latter property by the form factor $\bm{f}$, and under the assumption of the first Born approximation, it is $ARS\propto \left|\bm{f}\right|^2 S$.\cite{born_wolf1999,Zemb2002}
However, either of these two quantities usually determines the angular scattering profile of established scattering interfaces almost completely.
The strong spatial correlations present in a periodic array cause the structure factor to compact into a Dirac delta distribution. 
Hence, scattering from periodic arrays is only allowed into a small fraction of all available angular directions, the well-defined diffraction orders, while any other direction is prohibited as $S$ is zero otherwise. 
A fully random arrangement of scatterers is here in stark contrast. 
There, no particular direction is enhanced nor suppressed due to the arrangement of the scatterers and the scattering response of the array essentially becomes identical to that of the individual scatterer.

Between these two diametrically opposed types of spatial configurations lies the parameter space of correlated disorder, which received increased interest from the photonics community in recent years.\cite{vynck2012,wiersmas2013,rockstuhlAmorphousNanophotonics2013,Riboli2014,Yu2020} 
On the one hand, a spectrally broadband operation and robustness in fabrication makes the use of disorder attractive.
On the other hand, structural correlations enable effective means for directional and spectral enhancement or suppression of scattering. 
A special class of disorder is the \textit{hyperuniform disordered} (HuD) case for which large-scale density fluctuations vanish as they do in periodic systems but a disordered and isotropic spatial configuration is maintained.\cite{torquato2018,Yu2020} 
HuD structures have recently seen increasing interest in applications and experimental realizations due to their unique properties such as isotropic band gap formation\cite{torquato2018,Yu2020,sun2018,sapienza2017,leseur2016,florescu2009,Man2013,Florescu2013,Milosevic2019}.

In this work, we will show how a hyperuniform configuration of scatterers enables a new leverage to tailor light scattering by enabling both $S$ and $\bm{f}$ to significantly shape the resulting scattering response of the array.
The lack of large-scale density fluctuations %of the hyperuniform arrangement 
represents a hidden symmetry and, just as long-range translational symmetry in periodic systems, leads to a strong suppression of small-angle scattering. 
However, the disorder present in our structures on short length-scales relaxes the bounds of $S$ for larger scattering angles up to the extend of random structures, thus leaving the composition of the scattering pattern to $\bm{f}$.

As scattering elements we choose TiO$_2$ nanodisks.
Low-loss high refractive index nanoparticles are renowned for their large scattering cross sections, thus constitute ideal scattering elements for this study\cite{Kuznetsovaag2016}. 
Furthermore, high-index nanoparticles have received considerable attention from the metamaterials community in recent years as they allow, e.g., the study of magnetic in addition to electrical resonances at optical frequencies.\cite{Decker2013,Kuznetsovaag2016,Kruk2017}
These Mie-type resonances give rise to novel ways for near and far field manipulation such as nonlinear frequency generation\cite{Kruk2017,Koshelev2020} and wavefront shaping\cite{Jang2018,Wang2018}. 
Nanodisks have been investigated in particular due to their rather simple fabrication while allowing good control over spectral features via the geometry of the disks.   
Based on findings of our previous work\cite{piechulla2018}, in addition to the above we report on a novel scalable method to fabricate substrates covered with TiO$_2$ nanodisk arrays of tailored nearly hyperuniform disorder by a self-organized nanosphere deposition process.

\section{Sample fabrication scheme}
The procedure to fabricate large-scale TiO$_2$ arrays of tailored disorder relies on a nanosphere deposition technique that we have published elsewhere\cite{piechulla2018}.
By exploiting the nanospheres as a template in a subsequent dry-etching process, we significantly extend our approach by the ability to obtain disk-shaped nanoparticles. 
The developed process scheme allows controlled etching of TiO$_2$, which is a material of particular interest to the photonics community due to its high refractive index in the visible spectral range, and is, to the best of our knowledge, here reported on for the first time.

\begin{figure}[t]
    \includegraphics[width=1.0\textwidth]{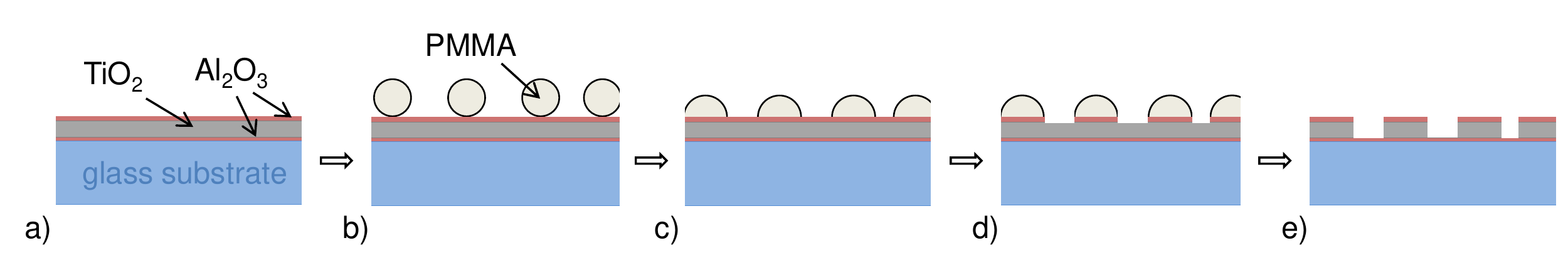}
    \caption{Fabrication scheme of disordered TiO$_2$ nanodisk arrays: (a) The substrate is functionalized with a Al$_2$O$_3$/TiO$_2$/Al$_2$O$_3$ layer stack. (b) Then a self-organized nanosphere deposition process is applied and the substrate is (c) tempered. The resulting domes serve as a template for the following (d) 2-step dry-etching process after which (e) arrays of TiO$_2$ nanodisks are obtained.}
    \label{fig:process}
\end{figure}

The fabrication scheme (see Figure~\ref{fig:process}) starts with standard microscopic slides that are functionalized with a Al$_2$O$_3$/TiO$_2$/Al$_2$O$_3$ multilayer with thickness 13.8\,nm, 231.0\,nm and 19.5\,nm, respectively.
We use atomic layer deposition (ALD) since this method allows great control over film thickness and produces dense layers.
The thin top layer serves a double purpose. 
Firstly, the surface charges of the Al$_2$O$_3$ film are essential for immobilizing the nanospheres in the subsequent nanosphere deposition. 
Second, this layer transforms into a hard mask imprinted by the nanospheres.
The thick TiO$_2$ layer provides the material of the nanodisks. 
The thin bottom layer serves as an etch stop to prevent overetching and damage to the substrate.

In the next step we apply our nanospheres deposition using PMMA nanospheres (diameter $D=499\pm 10$\,nm) in aqueous dispersion. 
In short, this technique exploits the electrostatic interactions of surface-charged nanospheres with each other and with the top Al$_2$O$_3$ layer.
The nanospheres (negatively charged) are driven towards the substrate (positively charged) and are immobilized as soon as they adsorb on the substrate surface. 
The nanospheres repel each other, thus modifying the probability of the location of adsorption for another nanosphere. 
After a certain density of adsorbed nanospheres is reached, no more nanospheres will attach since the repelling forces due to already adsorbed nanospheres become dominant.
Therefore, our deposition process is not only self-organized but also self-limiting, which makes it very cost-effective. 
A prepared dispersion can be applied multiple times since only a tiny fraction of nanospheres is used up during one deposition.  
By precisely setting the ionic strength of the dispersion, here via the addition of KCl, screening of the nanosphere surface charges within the aqueous dispersion can be controlled, providing an effective lever to control the resulting nanosphere density on the substrate. 
Details and theoretical background of our nanosphere deposition method are presented elsewhere\cite{piechulla2018}.

After the nanosphere deposition, the substrate is rinsed in deionized water and directly transferred to ethylene glycol and heated to 144$^\circ$C to slightly soften the nanospheres and increase the particle-substrate bond.
This step prevents particle aggregation due to surface tension of the liquid film during drying.
After rinsing with water and drying in air, the samples undergo an additional tempering step (30\,min at 155$^\circ$C) to change the nanospheres' shape to a dome-like structure such that the area below a nanosphere is fully covered with PMMA (Figure~\ref{fig:process}c).

The transition from the nanosphere pattern to a Al$_2$O$_3$ hard mask is performed via reactive ion-etching (RIE).
As shown by Dekker {\it et al.}, Al$_2$O$_3$ prepared by ALD provides an excellent hard mask for dry etching processes due to its chemical stability and superior film quality even in thin layers.\cite{DekkerAl2O3mask}
Due to the low volatility of AlF$_3$, Al compounds can not be etched by fluorine-based dry chemistry without a significant physical component. 
However, they are readily etched in chlorine chemistry.\cite{Powell1984,KooAlox2005,YunAlox2008,DekkerAl2O3mask}. 
Based on this, we developed the hard mask patterning process (see methods section for technical details).
The plasma is ignited in a BCl$_3$/Ar atmosphere with a low DC bias to achieve a sufficient Al$_2$O$_3$ over PMMA etching selectivity of $\approx 0.4$  \cite{YunAlox2008}.
The etch rate of Al$_2$O$_3$ was found to be $\approx 1.1$\,nm/s and the etch time was chosen to be 29\,s, which results in slightly overetching the hard mask layer while keeping the PMMA nanospheres sufficiently intact, which are afterwards removed in a pure O$_2$ plasma.

Unlike Al$_2$O$_3$, TiO$_2$ is etched by fluorine chemistry. 
Based on the work of Choi {\it et al.},\cite{ChoiDryetchingproperties2013} a CF$_4$-O$_2$-based process has been developed resulting in a TiO$_2$ etch rate of $\approx 1.5$\,nm/s and a TiO$_2$ over Al$_2$O$_3$ selectivity of $\approx 29$.
Hence, the top hard mask layer was sufficiently thick. 
For the 231.0\,nm TiO$_2$ layer, we etched for 240\,s, i.e. longer than actually needed. 
However, the Al$_2$O$_3$ etch stop layer beneath limits the etch depth and thereby increases repeatability of the manufacturing process (Figure~\ref{fig:process}e).
The requirements of high selectivity, high aspect ratios and smooth surfaces are difficult to meet in fluorine-based TiO$_2$ etching compared to, e.g., silicon processes \cite{GarayInductivecoupleplasma2015,Jansenblacksiliconmethod1995}, as parameters for vertical sidewalls and good selectivity also lead to self-masking effects and, thus, rough surfaces.
In our case, etching down to the etch stop layer reduces the roughness to an acceptable level for our investigations and produced relatively smooth surfaces (see Figure~\ref{fgr:SEM}b) compared to previously published results\cite{HotovyDryetchingcharacteristics2014,AdzhriReactiveIonetching2015a}.
We note that reactive ion etching processes have been successfully applied in industrial process chains and therefore do not principally impede the scalability of our fabrication method \cite{piechullaIncreasedIonEnergies2011,volkHoneycombStructureMulticrystalline2015}.

\section{Results and discussion}
The key objective of this study is to combine scattering properties of the individual TiO$_2$ nanodisk (form factor $\bm{f}$) with those introduced by positional correlations of their disordered arrangement (structure factor $S$). 
Therefore, we choose the dimensions of the nanodisks, diameter $D$ and height $h$, and the characteristic distance $r_0$ of the array, i.e. the typical next-neighbor center-to-center distance defined as the first maximum of the pair correlation function $g_2(r)$ of the array\cite{torquato2018}, such that the resulting spectral features overlap. 
Given the refractive index of our amorphous TiO$_2$, $n\approx$ 2.54 - 2.31 for a wavelength range of $\lambda=450-1000\,\text{nm}$, preliminary calculations using the finite-element method (FEM) predict a number of multipolar resonances in the visible and near-infrared spectral region for individual nanodisks of about 230\,nm height and 500\,nm diameter.
Accordingly, the nearest available particles size from stock, $D=499$\,nm, was chosen for the template fabrication.
For the structural resonances to spectrally coincide with the resonances of the individual nanodisks, $r_0$ needs to be of the same order as the wavelength.  
We have previously demonstrated how $r_0$ can be varied by the deposition conditions. 
Furthermore, we are able to statistically predict particle patterns based on a modified random sequential adsorption (RSA) model of soft spheres near the saturation density.\cite{piechulla2018}
Thus, we experimentally fabricated a set of samples comprising glass substrates covered with disordered TiO$_2$ nanodisk patterns with desired geometrical dimensions. 
Additionally, one pattern that ideally shows no structural correlations, i.e. of random disorder, was fabricated by exposing the functionalized substrate to the nanosphere dispersion for only a short period of time such that the particle density is far from saturation. 
In the following we discuss the properties of such samples first from a structural and later from an optical perspective.

\subsection{Structural properties}

\begin{figure}[t]
    \includegraphics[width=1.0\textwidth]{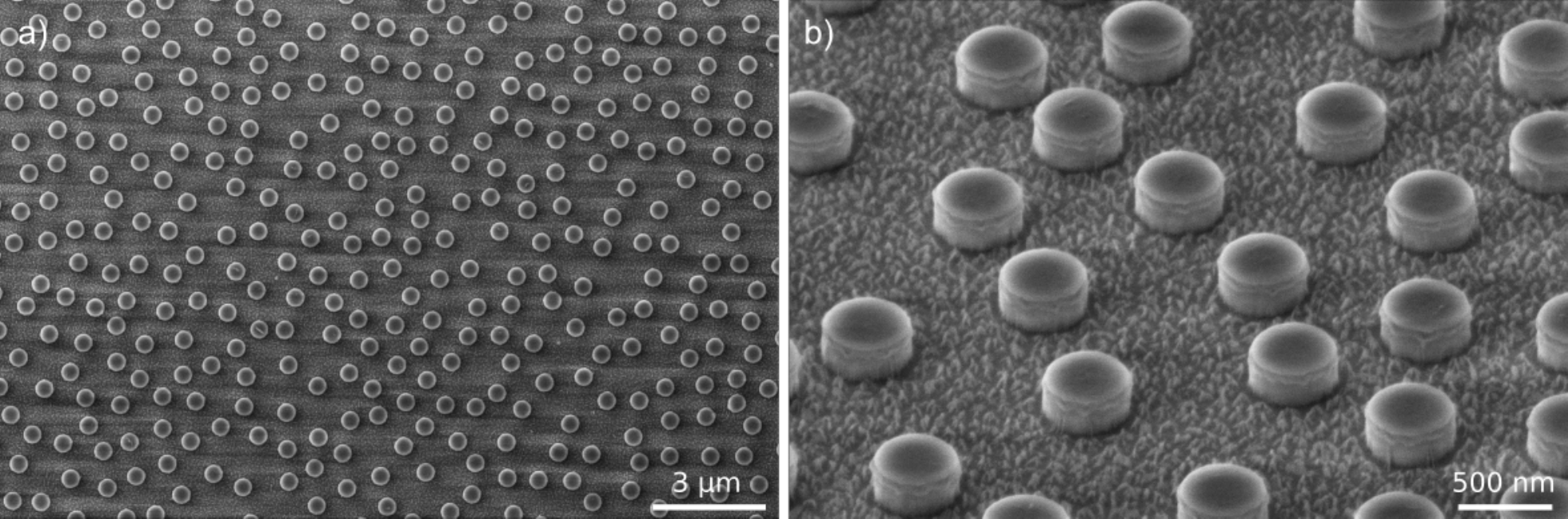}
    \caption{Scanning electron micrographs of a fabricated TiO$_2$ nanodisk array on glass substrate in a) top view and b) viewed at 52$^\circ$ inclination from normal. The disk density of this sample is 1.15 $\mu$m$^{-2}$ and the characteristic center-to-center distance is $r_0=828\,$nm.
    }
    \label{fgr:SEM}
\end{figure}

We successfully fabricated disordered arrangements of TiO$_2$ nanodisks.
Exemplary images of the fabricated TiO$_2$ nanodisk structures are shown in Figure~\ref{fgr:SEM}. 
We determine an average nanodisk diameter of 455.0\,$\pm$\,5.4\,nm and an average height of 231.0\,nm. 
We attribute the nanodisk diameter dispersion to the size dispersion of the nanospheres used to prepare the mask template. 
In comparison to the nanospheres (499\,nm diameter), the disks are around 10\,\% smaller in diameter, which is a result of the tempering process prior to pattern transfer.
The error on disk height is negligible due to the excellent control of ALD deposition over layer thickness.

\begin{figure}[t]
    \includegraphics[width=1.0\textwidth]{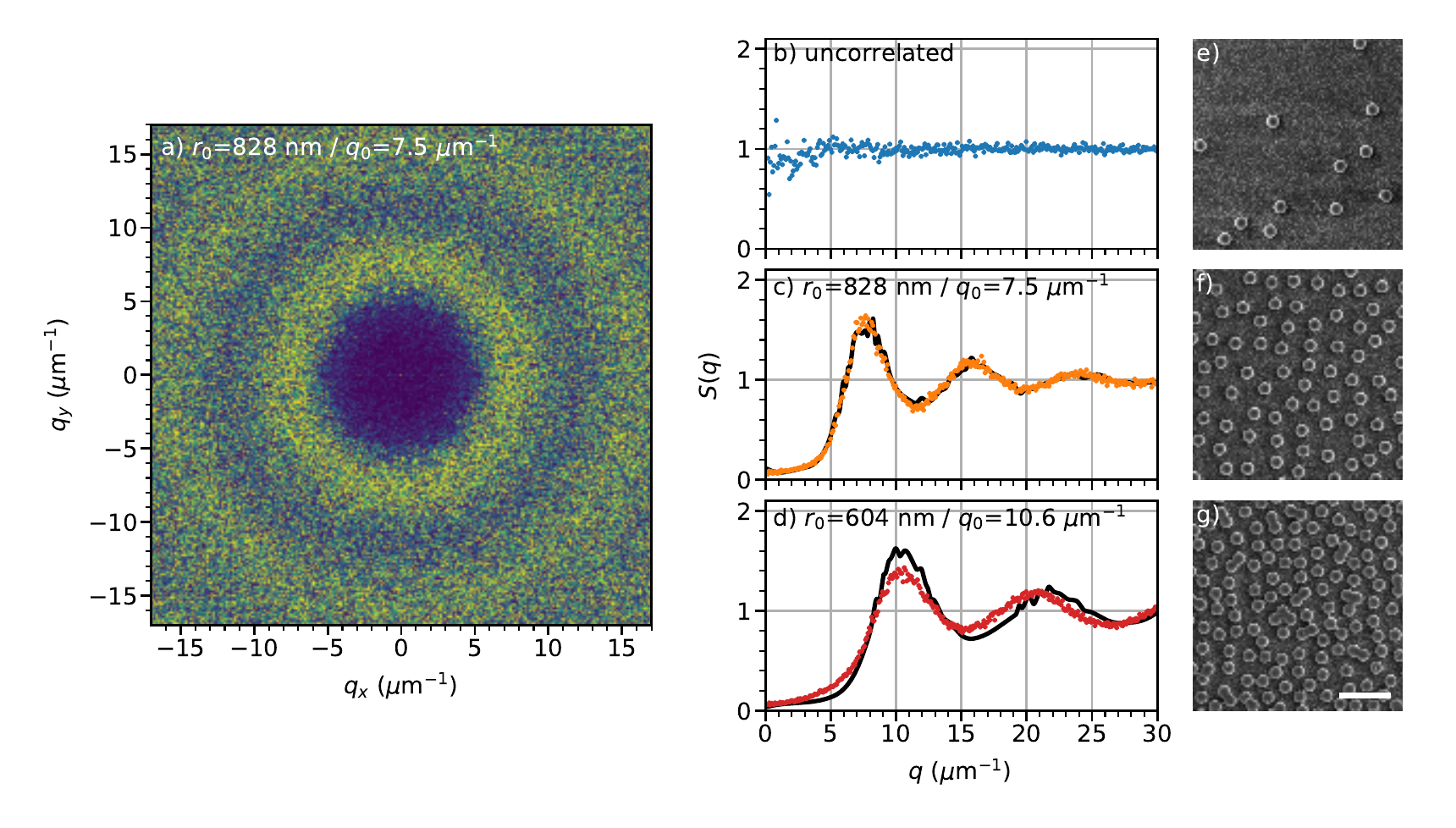}
    \caption{a)  Typical structure factor $S(\bm{q})$ for a fabricated disordered pattern. 
    b)-d) Angular average $S(q)$ for the uncorrelated pattern and two correlated patterns from experiment (dots) and simulations (solid black lines). 
    e)-g) Exemplary electron micrographs (detail) corresponding to structure factors to their left (scale bar: 2\,$\mu$m).}
    \label{fgr:Sq}
\end{figure}

The disks are homogeneously distributed across the entire substrate surface in a disordered pattern  (see characterization methods for pattern extraction)
The statistics of our patterns can be appropriately captured in Fourier space.
We consider the size dispersion of the disks sufficiently small to consider the structure factor instead of the spectral density.
However, the latter would be a more appropriate quantity to characterize polydisperse particle patterns \cite{dreyfus_diagnosing_2015}.
The (static) structure factor of a $N$-point pattern $\bm{r}_i$ is \cite{liu_direct_2016, allen1999}
\begin{equation}
S(\bm{q})=\frac{1}{N}\sum\limits_{i,j}^{N} e^{-i\bm{q}(\bm{r}_{i}-\bm{r}_{j})}
\label{eqn:Sq}
\end{equation}
with wave vectors $\bm{q}$. 
In this work, we write $\bm{q}$ for wave vectors that lie in the plane of the nanodisk array and $\bm{k}$ for wave vectors of propagating free-space light modes which in general have a component perpendicular to the array plane.
The point patterns are isotropic, thus $S$ is rotationally symmetric, exemplary shown in Figure~\ref{fgr:Sq}a.
The angular average $S(q)=S(|\bm{q}|)$ is plotted for typical samples in Figure~\ref{fgr:Sq}b-d together with $S(q)$ of predicted patterns.

Figure~\ref{fgr:Sq}b shows $S(q)$ for the sample on which the nanosphere deposition was stopped prior saturation.
As a result, spatial correlations are negligible and $S(q)\approx1$ for all $q$ and we will refer to this pattern in the following as \textit{uncorrelated}.
The deviation from unity near $q=0$ stem from the fact that particles maintain a minimum distance, which becomes relevant at finite densities.
However, the random nature of the process prevails.

The particle patterns of the samples corresponding to Figures~\ref{fgr:Sq}c and d are saturated and reveal strong correlations, i.e. $S(q)$ clearly deviates from unity.
Each structure factor $S(q)$ of these patterns holds a characteristic peak $S_\text{max}=S(q_0)$ at around $q_0=7.5\,\mu$m$^{-1}$ and $q_0=10.6\, \mu$m$^{-1}$, corresponding to a characteristic next-neighbor distance of about $r_0=828$\,nm and $r_0=604$\,nm, respectively, that is also indicated in the figure.
For the sparsely coated saturated sample (Figure~\ref{fgr:Sq}c,f), we find excellent agreement between our modified RSA model \cite{piechulla2018} and experiment, whereas for the densely coated sample (Figure~\ref{fgr:Sq}d,g) a lower value of $S_\text{max}$ and a slight shift of about $-0.1\,\mu$m$^{-1}$ compared to the prediction can be observed.
A possible explanation for this deviation in the denser sample is the occurrence of aggregates, i.e. touching particles, which is not accounted for in our deposition model. 
To experimentally achieve a certain characteristic distance $r_0$, we carefully adjust the electrostatic potential landscape in which the nanospheres move through during the deposition process and a rather shallow potential minimum (in the order of $-k_\text{B}T$) gives rise to a preferred next neighbor distance.\cite{piechulla2018}
However, as soon as two spheres touch each other, irreversible aggregation occurs.
Thus, the smaller the characteristic distance aimed for, the higher the probability of aggregate formation, which is supported by our observations in numerous other experiments.
Furthermore, aggregates would also explain increased values of $S(q)$ for $0<q<q_0$ in comparison to our model, as aggregate formation occurs in a random Poisson-like manner and, therefore, enhances long-range particle density fluctuations.
Therefore, aggregation and its effect on the resulting structure factor should be taken into account when considering high particle densities.
For the samples used in this work, the number share of particles present in configurations of two or more touching particles was found to be 2.6\,\%, 3.0\,\%, 9.5\,\%, and 31.2\,\% for $r_0=828$\,nm, 713\,nm, 649\,nm, and 604\,nm, respectively. %$r_0=713$\,nm, $r_0=649$\,nm, and $r_0=604$\,nm, respectively.
Due to the strong increase in aggregation for $r_0=604$\,nm, the pattern may be dubbed over-saturated.
Nevertheless, the general shape of the structure factor is still preserved (Figure~\ref{fgr:Sq}d).
Furthermore, we note that the final structure was obtained by patterning a flat layer. 
Therefore, all nanosphere aggregates result in in-plane nanodisk aggregates, i.e. a binary height profile with no component in the third dimension.

For our samples of correlated disorder, $S(q)$ becomes small for $q$ approaching zero, which is equivalent to a suppression of density fluctuations on large length scales.
When ${S(q\rightarrow0)=0}$, structures are referred to as \textit{hyperuniform}, and \textit{stealthy hyperuniform} when $S(q)$ vanishes for an entire range of $q$ around the origin.\cite{torquato2003, torquato2018}
In real world structures, the strict criterion of $S(q\rightarrow 0)=0$ can obviously not be achieved due to fundamental aspects such as finite size and temperature effects to the least. 
A practical criterion that has recently been proposed is the hyperuniformity metric\cite{kim_effect_2018,torquato2018}
\begin{equation}
H=\frac{S(q\rightarrow 0)}{S_\text{max}}.
\label{eqn:H}
\end{equation}
A system is considered as \textit{effectively} hyperuniform if $H$ is of the order of $10^{-4}$\cite{kim_effect_2018}, which is stricter than the requirement for \textit{nearly} hyperuniform patterns with $H$ on the order of $10^{-2}$\cite{torquatoRSA2006}.  
In our case, $H\approx 0.032$ and the lower limit of the structure factor is $S(q\rightarrow 0)\approx 0.05$, which is slightly lower than values found for 2D random sequential adsorption of hard sphere patterns\cite{zhangPreciseAlgorithmGenerate2013}.
Therefore, we consider the structures shown here nearly hyperuniform disordered.

Besides fundamental reasons, we suspect further limitations similar to those  applying to conventional RSA patterns hindering the achievement of lower degrees of hyperuniformity. 
In RSA patterns, a strictly sequential adsorption of particles onto the substrate surface is assumed and except the condition that particles must not overlap no interaction potential is accounted for.
Thus, the precise position of an adsorbed particle carries a rather large amount of randomness leading to density fluctuations at large length scales, i.e. $S(q\rightarrow 0)>0$.
In our case, the particles interact with each other via repulsive electrostatic and attractive van-der-Waals forces\cite{piechulla2018}.
The probability for a particle to adhere in the vicinity of already attached particles with a characteristic distance -which we are able to set experimentally- is therefore modulated and leads to increased correlations of particle positions.
Additionally, the adsorption of particles is not strictly sequential so there might be some degree of coordination while particles are mobile.
However, the particle ensemble cannot achieve a collective configuration to minimize potential energy since they are immobilized as soon as they touch the substrate.
Thus, density fluctuations are not fully suppressed and it can therefore be assumed that $H$ in our nanosphere deposition will be at least identical or, as observed here, smaller to the RSA case.
At this point, the limits of $H$ in self-organized bottom-up techniques like ours are still an open question.
Nevertheless, the relative stealthiness of our patterns, i.e. $S(q)$ being rather small for a range of $q$'s around the origin, has a strong impact on their scattering properties as we will show in the next section.

\subsection{Optical properties}

\subsubsection{Numerical method}

\begin{figure}
    \includegraphics[width=0.3\textwidth]{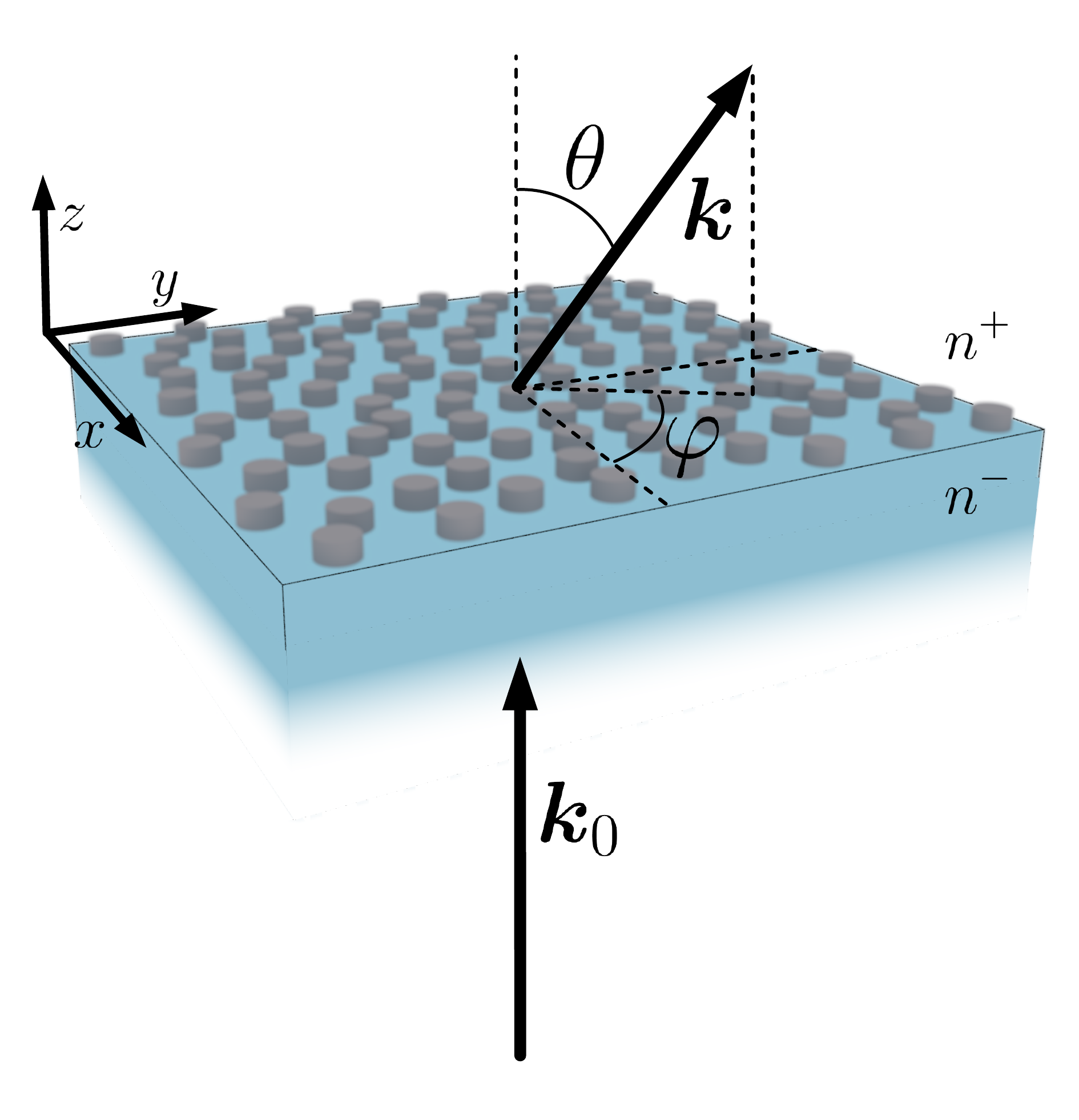}
    \caption{Sketch of the notation used throughout this work.
    The nanodisk array lies in the $xy$-plane on a glass substrate and the incoming light is travelling through the substrate at normal incidence with regard to the array plane.
    }
    \label{fgr:notation}
\end{figure}

To numerically reproduce the optical response of our fabricated TiO$_2$ nanodisk arrays, we apply the first Born approximation,\cite{born_wolf1999} i.e. the exciting light field for each scatterer is assumed to be the unperturbed incoming light field $\bm{E}_\text{in}$. 
Throughout this work we consider plane wave irradiation through a transparent substrate of refractive index $n^- = 1.52$ occupying the $z \leq 0$ half space.
The $z > 0$ half space is vacuum with $n^+ = 1$, see Figure\,\ref{fgr:notation} for notation.
The plane wave with field amplitude $\bm{E_0}$ is $x$-polarized and normally incident onto an infinite array of identical scatterers covering the substrate, i.e. $\bm{E}_\text{in}(\bm{r}) = E_0\hat{\bm{x}}\, e^{i(k_0 z-\omega t)}$.

The scattered far field ($kr \gg 1$) of an individual scatterer can be expressed as an outgoing spherical wave (we omit the harmonic time dependence $e^{-i\omega t}$ in the following)
\begin{equation}\label{sph_wav}
    \bm{E}_\text{s}(\bm{r}) = E_0 \frac{e^{ i\bm{k}\cdot\bm{r} }}{r}  \bm{f}.
\end{equation}
$\bm{f}=\bm{f}(\hat{\bm{E}}_0,\bm{k}_0,\bm{k})$ is the \textit{form factor} and describes the normalized scattered field amplitude of an individual scatterer for a given polarisation $\hat{\bm{E}}_0=\bm{E}_0/|\bm{E}_0|$ and incoming and scattered wave vectors $\bm{k}_0$ and $\bm{k}$, respectively.
Furthermore, $|\bm{f}|^2=\frac{\text{d}\sigma}{\text{d}\Omega}$ is the differential scattering cross section or radiation pattern of the individual particle.
$\bm{f}$ can be readily obtained by various means, e.g. analytically by Mie theory in case of a spherical particle, or by rigorously solving Maxwell's equations with techniques such as FEM. 
We write $\bm{f}_0$ for $x$-polarized irradiation normal to the array plane, i.e. $\bm{f}_0(\bm{k})=\bm{f}(\hat{\bm{x}},k\hat{\bm{z}},\bm{k})$.
Using the angular spectrum representation of Equation\,\ref{sph_wav} in its asymptotic limit ($kr\rightarrow\infty$),\cite{wolf1995} we find the normal incidence transmission and reflection of an infinite array of identical scatterers in terms of the form factor to be 
\begin{eqnarray}
 \label{T} T & =  \frac{n^+}{n^-} & \left|  t\hat{\bm{E}}_0 + \rho \frac{2\pi}{ik^+}\bm{f}_0^+\right|^2 \\
 \label{R} R & =  & \left|  r\hat{\bm{E}}_0 + \rho \frac{2\pi}{ik^-}\bm{f}_0^-\right|^2
\end{eqnarray}
with areal density of scatterers $\rho$, Fresnel's transmission and reflection coefficients $t$ and $r$ of the bare substrate interface, wave numbers $k^+$ and $k^-$ in 
substrate and vacuum, and $\bm{f}_0^\pm = \bm{f}_0 (\bm{k}=\pm\bm{k}_0)$, respectively.

Scattering into the directions normal to the array plane, $\bm{k}=\pm k\hat{\bm{z}}$, %($\theta=0^\circ$ and $\theta=180^\circ$) 
depends on $\bm{f}_0$ and $\rho$, but not on the particular spatial arrangement of scatterers.
Interference of these fields with $t\hat{\bm{E}}_0$ and $r\hat{\bm{E}}_0$ (Equations\,\ref{T} and \ref{R}) determines the normalized power $P_\text{sca}$ that undergoes off-normal scattering, %from energy conversation it follows that 
$P_\text{sca}=1-T-R$ (neglecting absorption). 
However, for off-normal scattering structural phase relations have to be accounted for by modulating %the radiation pattern of the individual particle,
$\left|\bm{f_0}\right|^2$ by the structure factor $S(q=k_\parallel)$, with the projection $k_\parallel$ of the scattered wave vector $\bm{k}$ onto the array plane.\cite{Zemb2002}
Using $P_\text{sca}=\int_\Omega ARS\, \text{d}\Omega$, with integration of the angle-resolved scattering $ASR$ over the unit sphere $\Omega$ except for $\text{d}\Omega = \text{d}\Omega(\bm{k} \pm k\hat{\bm{z}})$, we can now write down the $ARS$ response of the array at normal incidence as
\begin{equation}\label{Born1}
    ARS
    = \frac{P_\text{sca}}{\int_\Omega\left|\bm{f_0}\right|^2 S\,\text{d}\Omega} \cdot \left|\bm{f_0}(\bm{k})\right|^2 S(k_\parallel)
    \hspace{1cm} \bm{k}\neq \pm k\hat{\bm{z}\, }. 
\end{equation}

The numerical approach described here significantly simplifies the computational efforts to predict the optical response of our structures.
The structure factor $S$ in Equation~\ref{Born1} can be taken either from experiment, e.g. by evaluating SEM pictures, or, if sufficient knowledge of the fabrication process is available, from theoretical modelling such as our modified RSA approach.\cite{piechulla2018}
Instead of numerically solving Maxwell's equations for a computational domain large enough to sufficiently approximate the statistics of a disordered array of scatterers, only the scattering response of the individual scatterer is calculated rigorously. 
We shortly discuss the applicability of the first Born approximation to our structures at the end of the discussion, see below.

\begin{comment}
In the first Born approximation (Equation~\ref{Born1}), near-field coupling between neighboring particles and multiple scattering are neglected.
The near-fields of high index nanodisk resonances are mostly confined to the inside, thus, coupling only occurs for particles very close to each other. 
It has been shown for Si nanoparticles with resonances in the visible that substantial coupling starts for surface-to-surface distances smaller than 25\,nm to 100\,nm.\cite{VandeGroep2016a}
Here, smallest typical surface-to-surface distances are 149\,nm, hence we conclude that the assumption of negligible near-field interaction is justified.
Multiple scattering can be neglected as we consider out-of-plane scattering of a nanodisks array at normal incidence.
\end{comment}

\begin{comment}
\begin{figure}
    \includegraphics[width=1.0\textwidth]{scatterFotos.png}
    \caption{Photographed forward scattering of TiO$_2$ nanodisks arranged in an (a) uncorrelated and (b) correlated pattern illuminated with collimated laser light ($\lambda=520$\,nm, normal incidence, unpolarized) onto a screen.
    (c) The observed ring-shaped pattern (thick green line) evolves from light waves scattered by each individual disk (thin green line) interfering constructively or destructively as prescribed by the structure factor (orange line) of the disks' spatial arrangement (Equation~\ref{eqn:scatter}).}
    \label{fgr:scatfoto}
\end{figure}
\end{comment}

\subsubsection{Angle-resolved scattering response}

\begin{figure}[t]
    \includegraphics[width=1.0\textwidth]{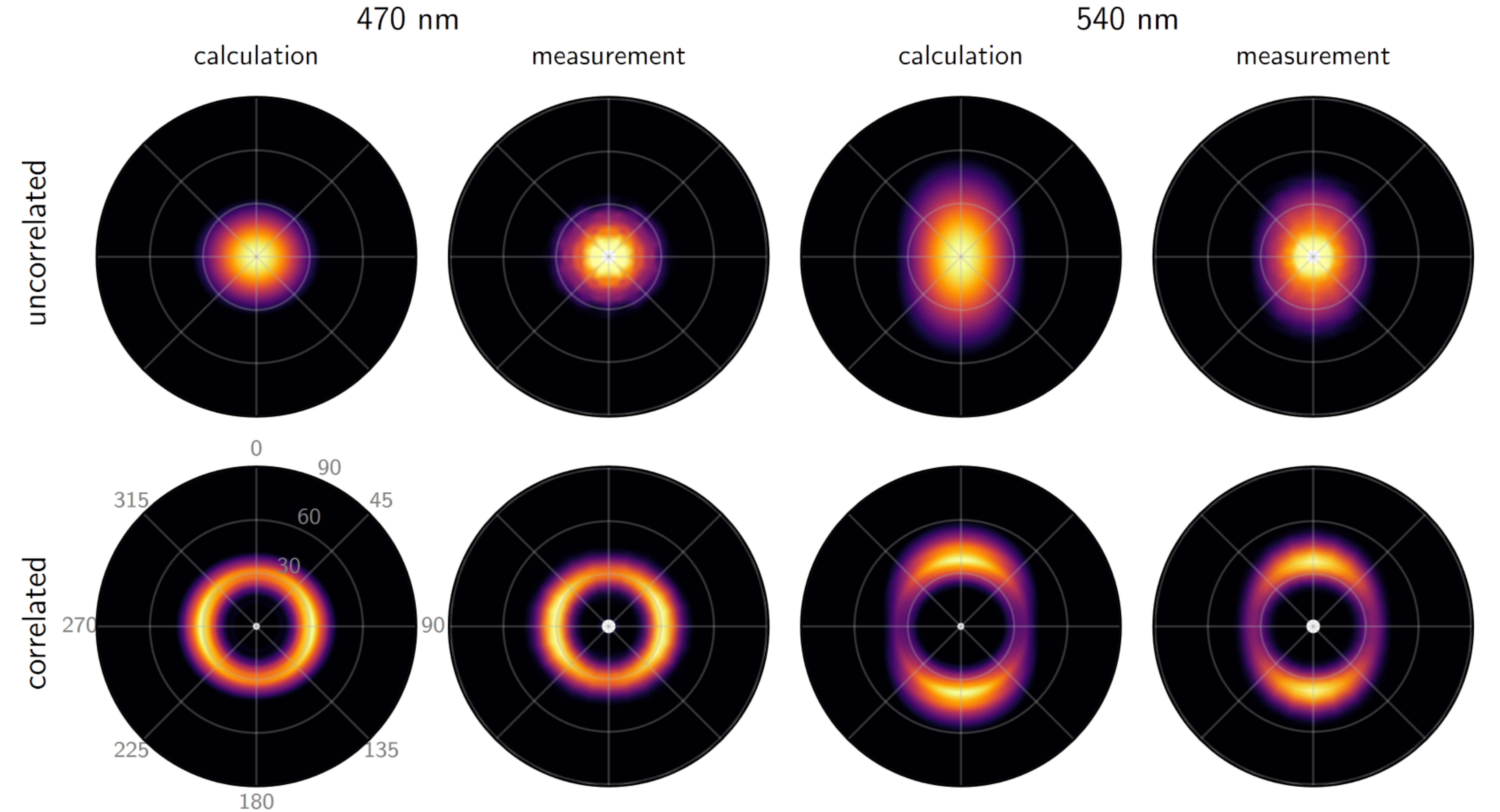}
    \caption{Measured and calculated $ARS(\theta,\varphi)$ in forward direction of an uncorrelated (top row) and nearly HuD array of nanodisks (bottom row). $S$ and SEM of the uncorrelated and correlated sample are shown in Figure\,\ref{fgr:Sq}b,e and c,d, respectively.
    The incoming light is $x$-polarized.
   }
    \label{fgr:polar}
\end{figure}

Figure~\ref{fgr:polar} illustrates how arranging identical TiO$_2$ nanodisks into a nearly HuD pattern has a remarkable impact on the resulting angular scattering response. % (see characterization methods details on measurement).
For two different wavelengths we compare the $ARS$ of the uncorrelated pattern in the top row and a nearly HuD pattern (characteristic distance $r_0=828$\,nm) in the bottom row. 
Additionally, corresponding numerical results after the approach described above are plotted using $\bm{f_0}$ calculated by FEM and $S$ taken from our modified RSA modelling algorithm.

For the uncorrelated sample, it is $S\approx1$ for all $q$'s (see Figure\,\ref{fgr:Sq}b), thus its $ARS$ essentially remains that of the individual nanodisk but scaled by the number of illuminated particles.
Scattering is strongest at $\theta = 0^\circ$ and quickly decreases as $\theta$ increases.
In contrast to this, scattering is very different for the nearly HuD pattern with a maximum at significantly larger angles and being mostly confined within a ring-shaped area.
Spatial correlations in these structures, e.g. expressed through the existence of the characteristic distance of neighbors $r_0$, leads to similar optical phase differences for off-normal scattered light waves.
Thus, destructive as well as constructive interference occurs and even though translational symmetry is missing, intensity maxima (ring-shaped area) or minima (angular area inside ring) appear well pronounced.
Opposed to the well-know diffraction of an optical grating, which is based on interference as well, constructive interference occupies a much larger solid angle due to a range of similar but not exactly equal distances between scatterers.
E.g., for ${\lambda=540}$\,nm we find that $\int_{\theta}\int_{\varphi=0}^{2\pi}ARS\,\text{d}\Omega=80$\,\% of forward scattering is confined within $\theta < 48^\circ$ for the uncorrelated case, while the same amount is confined within $25^\circ<\theta<56^\circ$ for the nearly HuD case.

In contrast to irradiation with unpolarized light, we here observe experimentally as well as numerically a $\varphi$-dependency of the scattering response, which is very subtle for most part of the investigated spectral range but is slightly visible at $\lambda = 470$\,nm and well observed for $\lambda = 540$\,nm.
In particular, for the nearly hyperuniform sample it can be seen that for $\lambda = 470$\,nm scattering reaches out further into the $y$- than into the $x$-direction, and even more pronounced for $\lambda = 540$\,nm but perpendicular to the former, i.e. reaching out further into the $x$- than into the $y$-direction.
The same anisotropy is also present for the uncorrelated sample, but less pronounced.

\begin{figure}[t]
    \includegraphics[width=\textwidth]{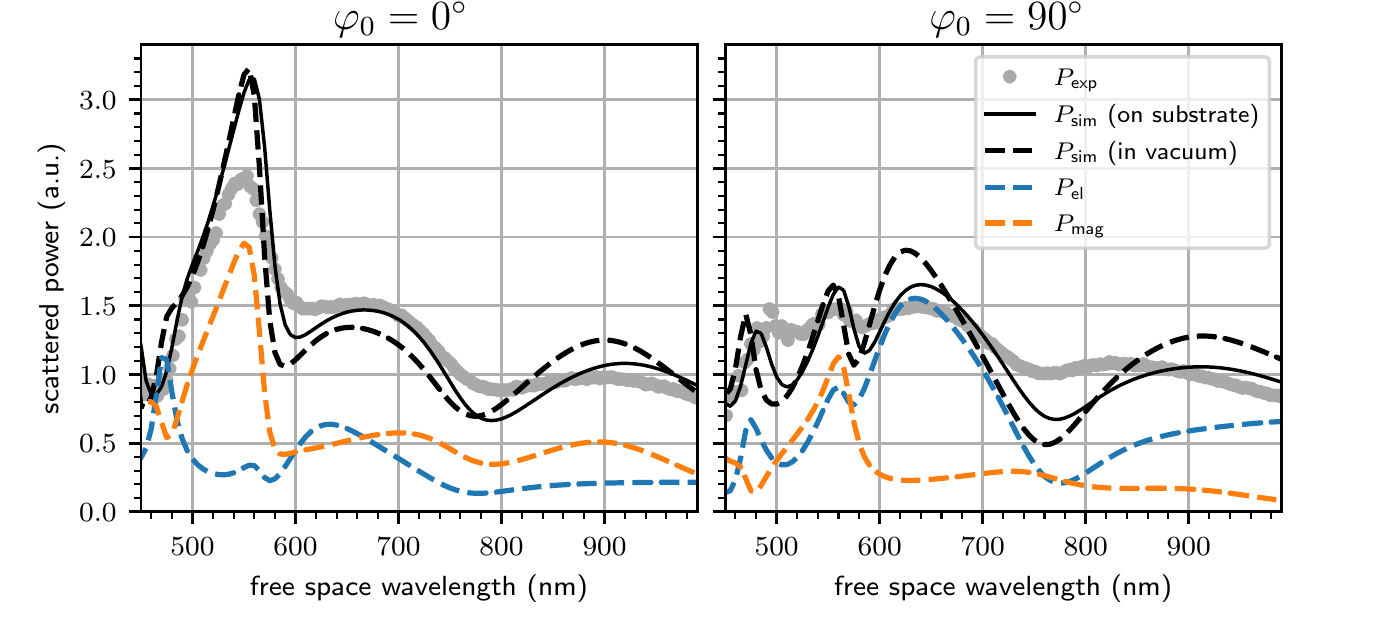}
    \caption{Forward scattering $P(\varphi_0,\lambda)$ into the plane defined by $\varphi_0 = 0^\circ$ ($xz$-plane) and $\varphi_0 =90^\circ$ ($yz$-plane).
    Experimental data of the uncorrelated sample together with simulations of a TiO$_2$ nanodisk on glass substrate as well as without substrate.
    Additionally, $P_\text{el}(\varphi_0,\lambda)$ and $P_\text{mag}(\varphi_0,\lambda)$ is plotted, i.e. the radiation power of the sum of electric and  magnetic dipole, quadrupole, and octupole, respectively, derived from multipole expansion analysis of the same nanodisk without substrate (see also \textit{supporting information}).
    }
    \label{fgr:mulipoles}
\end{figure}

To elucidate our observations, we look into the $\varphi$-dependency in more detail in Figure\,\ref{fgr:mulipoles}.
Plotted is the power scattered into the $xy$- and $yz$-plane in forward direction, $P(\varphi_0,\lambda)=\int_{\theta=0}^{\nicefrac{\pi}{2}}\int_{\varphi=\varphi_0-\Delta\nicefrac{\varphi}{2}}^{\varphi_0+\Delta\nicefrac{\varphi}{2}}ARS(\lambda)\,\text{d}\Omega$ with $\varphi_0=0^\circ$ and 90$^\circ$, respectively,  and $\Delta\varphi = 3.5^\circ$, and of the uncorrelated sample together with FEM simulations of a single TiO$_2$ nanodisk on glass substrate.
Furthermore, we performed a multipole expansion analysis of a single TiO$_2$ nanodisk. 
As multipole expansion is only defined for a scatterer embedded in a homogeneous medium, we analyse a TiO$_2$ nanodisk with identical geometrical dimensions in vacuum (dashed lines).
Comparing $P(\varphi_0,\lambda)$ for the isolated disc in vacuum with the case of the disk on glass substrate, we find only minor deviations, particularly for the spectral feature at $\lambda = 540$\,nm. 
This indicates that the removal of the substrate has only little impact on the optical properties of the disk and insights obtained from the multipole expansion considering the isolated disc can be applied to the substrate case. 

The multipole expansion reveals that for our chosen nanodisk dimensions a plethora of modes ranging from magnetic and electric dipoles up to octupoles contribute significantly to the scattering response withinin the spectral range of interest (see Figure S1 in \textit{supporting information}). None of the spectral features found in the scattering spectra can be accounted to a single mode but a mix of several modes.
However, comparing the power emitted by the sum of electric dipole, quadrupole, and octupole with the sum of their magnetic counterparts, $P_\text{el}(\varphi_0,\lambda)$ and $P_\text{mag}(\varphi_0,\lambda)$, respectively (colored dashed lines in Figure\,\ref{fgr:mulipoles}), reveals that scattering at $\lambda = 470$\,nm is predominately caused by electric modes, whereas at $\lambda = 540$\,nm it is predominately caused by magnetic modes.

The electric moments excited in the nanodisks share an equatorial plane that is oriented perpendicular to the one shared by the magnetic moments, according to the geometry of our experiment the $yz$- and the $xz$-plane, respectively.
Therefore, since their equatorial planes are also the planes of preferred scattering, excitation of these two kind of modes explains the observed $\varphi$-anisotropy observed in Figure~\ref{fgr:polar}.
However, as both kind of modes emit into small $\theta$s, the discussed effect is only significant at larger $\theta$s.
As scattering concentrates at rather small $\theta$s for the uncorrelated case, the dependency of the scattering pattern on the nature of the excited mode is only weak.
But arranging the same nanodisks into a nearly HuD pattern strongly emphasizes this effect.
Due to its structure factor, small-angle scattering is suppressed, hence giving an enhanced spatial contrast to differentiate between the nature of the excited modes.
To be specific, for TiO$_2$ disks arranged in an uncorrelated pattern the ratio $P(\varphi_0=0^\circ,\lambda)/P(\varphi_0=90^\circ,\lambda)$ at $\lambda=540$\,nm is 2.2, but it more than doubles to 5.2 in the nearly HuD case.

\subsubsection{Spectral dependence of scattering response}

\begin{figure}
    \includegraphics[width=.65\textwidth]{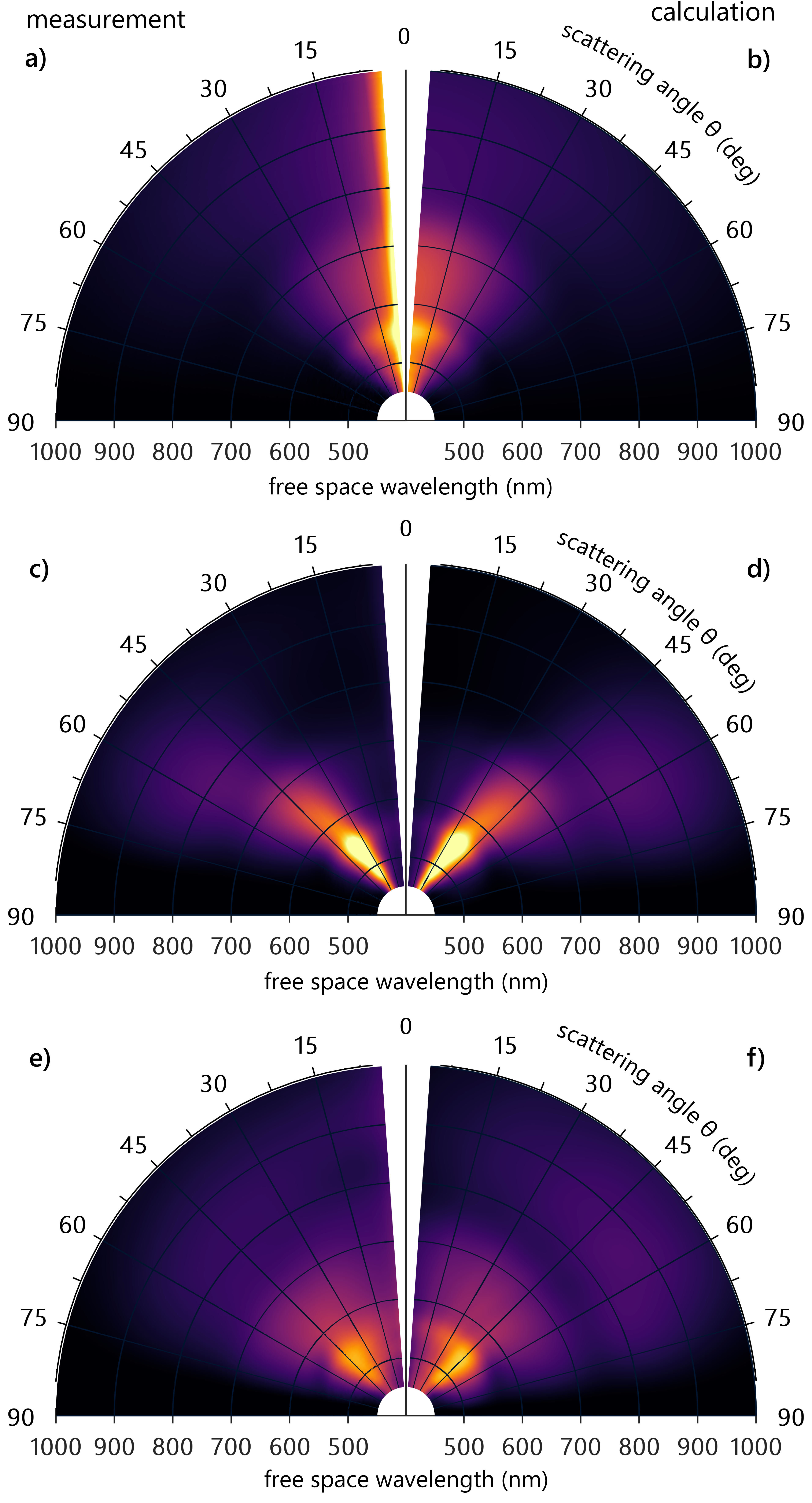}
    \caption{Measured (left column) and calculated (right column) $\varphi$-averaged angle-resolved scattering response $ARS(\theta,\lambda)$ in forward direction of (top row) the uncorrelated TiO$_2$ nanodisk array, and the nearly HuD array with (centre row) $r_0 = 828$\,nm, and (bottom row) $r_0 = 604$\,nm. 
    Color scaling is identical for (a) and (b) and (c)-(f), respectively.
    }
    \label{fgr:corr_vs_uncorr}
\end{figure}

In Figure~\ref{fgr:corr_vs_uncorr}, we capture the spectral dependence of scattering in near HuD TiO$_2$ nanodisk arrays by plotting the calculated and experimentally measured $ARS$ against wavelength $\lambda$ and forward scattering angle $\theta$.
As the $\varphi$-dependency was discussed in the previous section and for the sake of clarity, the values are averaged over $\varphi$. 
The top row shows the scattering response of the uncorrelated array, i.e. essentially the unmodified response of the individual TiO$_2$ nanodisk, the centre and bottom rows show the scattering response of two nearly HuD arrays with different characteristic distances $r_0 = 828$\,nm, and $r_0 = 604$\,nm, respectively. Both of them have been discussed in Figure~\ref{fgr:Sq}.
As indicated above for single wavelengths, experimental and calculated ARS agree as well for the whole considered spectral range.

The unmodified scattering response of the uncorrelated pattern is characterized by spectrally broadband forward scattering with its maximum at $\theta=0$. 
The strongest scattering response is around $\lambda=560$\,nm, which can be attributed to the simultaneous excitation of the magnetic dipole, quadrupole, and octupole mode (see also \textit{supporting information}, Figure \ref{si:multipoles}).
The center row of Figure~\ref{fgr:corr_vs_uncorr} shows the scattering response of the nearly HuD sample that was also under study in the previous section. 
In accordance to the observations for single wavelengths, the ARS is modified for the whole considered spectral range as well, i.e. transformed into a ring-shaped pattern with a scattering maximum $\theta \neq 0$.
Surprisingly, the angular range covered by the ring stays almost constant for a rather large spectral bandwidth. 
To be precise, for $\lambda = 450 - 750$\,nm the major part of scattered light stays within  $30^\circ < \theta < 50^\circ$, for $\lambda = 800 - 1000$\,nm, the ring radius rather abruptly increases to occupy an angular range of $50^\circ < \theta < 70^\circ$.

\begin{figure}[!t]
    \includegraphics[width=1\textwidth]{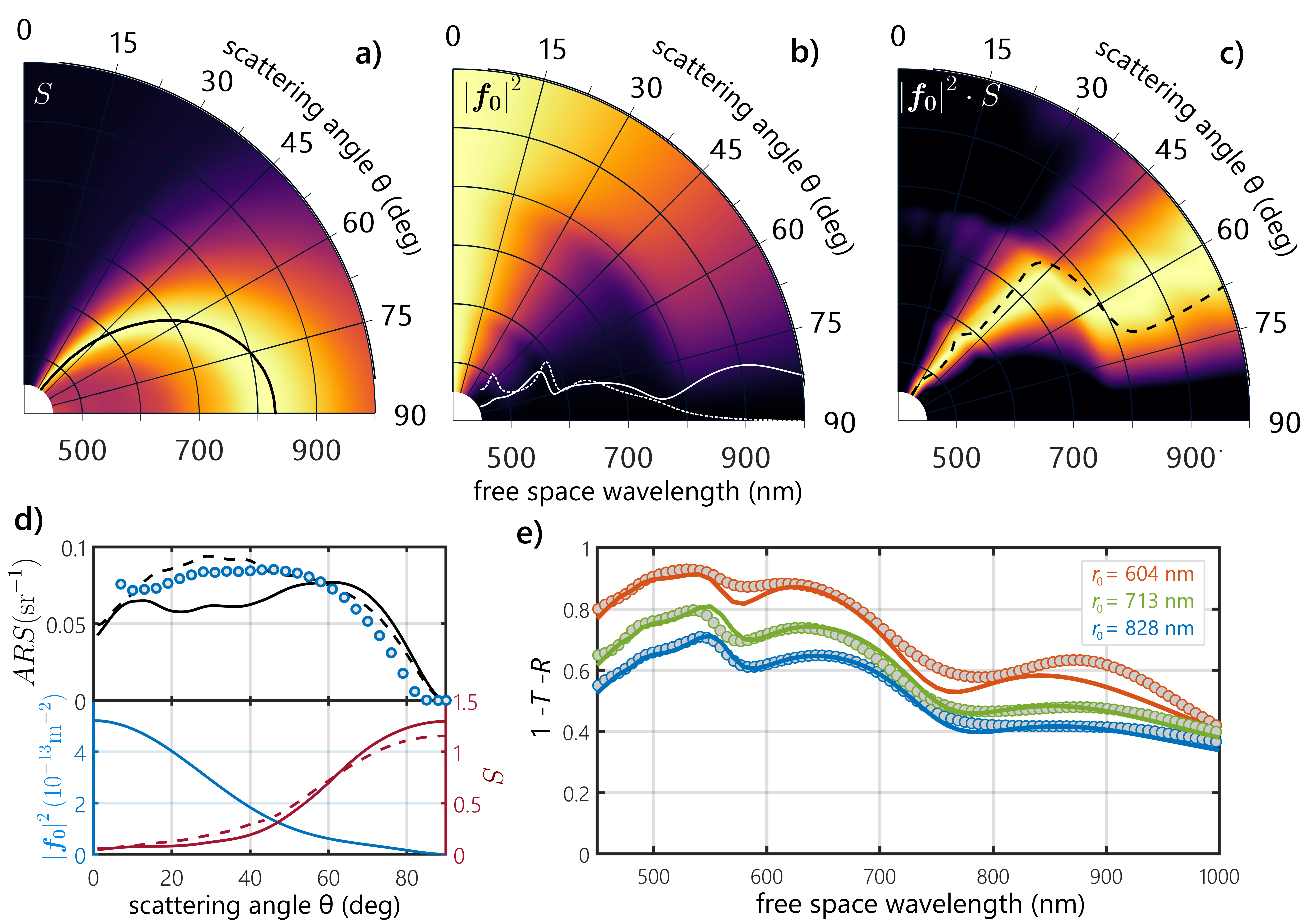}
    \caption{(a) $S(\lambda,\theta)$ of our nearly HuD pattern with $r_0 = 828$\,nm.
    (b) $\varphi$-averaged $\left|\bm{f}_0\right|^2$ normalized by its maximum value for each wavelength.
    The overlayed white line indicates the contribution of electric and magnetic dipole modes, $|a_1|^2+|b_1|^2$, the dashed line the contribution of higher order modes, $\sum_{i=2,3}|a_i|^2+|b_i|^2$. 
    (c) $ARS$ of the nearly HuD array normalized by its maximum value for each wavelength.
    The dashed black line indicates a contour line of the data shown in (b) at $\left|\bm{f}_0\right|^2 =0.35$.
    (d) Experimental (circles) and calculated (black) $ARS$, $\left|\bm{f}_0\right|^2$, and $S$ (all for $\lambda=700$\,nm) of the sample shown bottom in Figure\,\ref{fgr:corr_vs_uncorr}. Full and dashed lines of $ARS$ and $S$ indicate use of simulated and experimental array patterns, respectively. 
    (e) Experimental (circles) and calculated (lines) $1-T-R$ of different nearly HuD samples with characteristic next-neighbor distances $r_0$.}
    \label{fgr:S_and_FF}
\end{figure}

Scattering in both periodic and (nearly) HuD structures is based on structural phase-induced interference, therefore it is intuitive to compare the two. 
In Figure\,\ref{fgr:S_and_FF}a we plot the same $S$ as in Figure\,\ref{fgr:Sq}c but converted $q$ to the corresponding scattering angle $\theta$ for each wavelength $\lambda$.
Since $S$ is a property of the arrangement of scatterers within the array, it is invariant with regards to constant $q$'s, and, thus, its course for varying $\theta$ and $\lambda$ is determined through $\theta = \arcsin{(\nicefrac{\lambda \, q}{2 \pi} )}$ (assuming normal incidence). 
This general relation is fundamental for periodic structures, such as optical gratings, since it sets the only angular directions periodic structures are allowed to scatter into as long as $q$ exactly equals a multiple of reciprocal grating vectors.  
The black line in Figure\,\ref{fgr:S_and_FF}a follows $S_\text{max}=S(q_0)$ of the HuD array.
Considering a periodic pattern with periodicity $a=r_0$, this line would also represent its (first) diffraction order. 
The most distinct difference between $S$ of a periodic and a HuD structure is that the first consists of a set of Dirac delta functions while the latter is a continuous function.
This implies the existence of a cutoff wavelength $\lambda_\text{c}=a$ for the periodic structure, i.e. for wavelengths $\lambda>\lambda_\text{c}$, off-normal scattering is disabled since then all scattered waves become evanescent in the normal direction.
In contrast to this, the continuous nature of $S$ of a HuD structure leads to a pronounced spectral broadening of any features of $S$ as clearly visible in Figure\,\ref{fgr:S_and_FF}a.
For $\lambda_\text{c}>r_0$, i.e. when light with wavelengths longer than the one corresponding to $S_\text{max}$, non-vanishing values of $S$ still stretch over a considerable angular and spectral range enabling scattering.
For example, light waves corresponding to $S(q=6\,\mu\text{m}^{-1})=0.5\cdot S_\text{max}$ are able to propagate in free space up to $\lambda=1050$\,nm.
In case of a truly stealthy HuD structure, for which $S(q)=0$ for some finite $q<Q$, a cutoff wavelength exists as well, but in contrast to the periodic case, this wavelength can in principle be much longer than the characteristic next-neighbor distance of scatterers within the array.

However, by comparing Figure~\ref{fgr:corr_vs_uncorr}c/d and Figure\,\ref{fgr:S_and_FF}a, the $ARS$ cannot be traced back to the structure factor alone. 
Indeed, we find that the form factor $\bm{f}$ has to be taken into account as well. 
In Figure\,\ref{fgr:S_and_FF}b we plot the radiation pattern $\left|\bm{f}_0\right|^2$ of the individual nanodisk.
Since we focus on the angular dependence for now, $\left|\bm{f}_0\right|^2$ is normalized by its maximum value for each wavelength, i.e. by its value in forward direction ($\theta=0$).
In this representation, the angular dependencies of the different involved multipoles become visible.
For $\lambda<800$\,nm, scattering is confined roughly within $\theta< 45^\circ$ but abruptly widens up to $\theta=75^\circ$ for $\lambda>800$\,nm.
Multipole expansion analysis reveals that this is due to the electric and magnetic dipole modes being the dominant modes excited in our nanodisks for $\lambda>800$\,nm (see overlayed full and dashed white lines in Figure\,\ref{fgr:S_and_FF}b). 
The resulting scattering pattern of the individual nanodisk depends on the detailed interference of all excited multipoles\cite{Mie1908}.
In contrast to higher order modes, such as quadrupole and octupole modes, the radiation pattern of pure dipole modes does not hold a $\theta$-dependence, i.e. does not posses any radiation lobes.
Therefore, as long as dipole modes are dominant their radiation pattern is only little affected by interference with higher-order modes and they keep their angularly broad scattering.  
This is not the case for $\lambda<800$\,nm, as electric and magnetic quadrupoles as well as octupoles are excited, resulting in a scattering pattern that is confined to a smaller $\theta$-range.

The final $ARS$ of scatterers with form factor $\bm{f}$ arranged into an array with structure factor $S$ is proportional to their product, $\left|\bm{f}_0\right|^2 \cdot S$ (Equation\,\ref{Born1}), which we plot in Figure\,\ref{fgr:S_and_FF}c.
Essentially, this is the same data as shown in Figure~\ref{fgr:corr_vs_uncorr}d, but again we normalize to the maximum value at each wavelength to better visualize the resulting main direction of scattering.
Additionally, a contour line of the normalized radiation pattern of the individual nanodisk (Figure\,\ref{fgr:S_and_FF}c) is overlayed following the value of $\left|\bm{f}_0\right|^2$ at the absolute scattering maximum of this sample, ($\lambda=560$\,nm, $\theta=35^\circ$, see Figure~\ref{fgr:corr_vs_uncorr}d).
In this representation, it becomes visible that the angular scattering properties of our nearly HuD structures is indeed not only governed by $S$. 
Comparing the contour line and the course of maximum scattering (light-yellow regions in Figure\,\ref{fgr:S_and_FF}c) reveals that the angular dependence of the individual scatterer is clearly imprinted into the angular dependence of the array.

This is in stark contrast to diffraction in periodic structures.
In these, the impact of $\bm{f}$ is limited to the intensity of a diffraction order, but $\bm{f}$ never affects the direction of propagation.
However, our observations are as well in contrast to truly random structures, in which $\bm{f}$ alone determines the scattering pattern since in those structures is $S=1$.
In fact, instead either $S$ or $\bm{f}$ dominating the scattering response, tailored disorder structures such as our HuD structures enable a precise interplay between $\bm{f}$ and $S$.

An example of how delicate this interplay can be reveals another sample with the same TiO$_2$ nanodisks but different characteristic distance $r_0=604$\,nm.
Its experimentally measured $ARS$ is plotted in Figure~\ref{fgr:corr_vs_uncorr}e and we observe a rather uniform scattering response with regards to $\lambda$ and $\theta$.
With $S$ from our modified RSA modelling we were not able to theoretically reproduce the $ARS$ as good as it is the case for the other samples.
We found better agreement using the experimentally evaluated $S$ (deduced from SEM pictures), the resulting $ARS$ is shown in Figure~\ref{fgr:corr_vs_uncorr}e. 
This can also be observed in Figure\,\ref{fgr:S_and_FF}d, where we plot  $\left|\bm{f}_0\right|^2$, $S$, and the resulting $ARS$ for $\lambda=700$\,nm. 
It can be recognized how the slight deviations between simulated (full line) and experimental  (dashed) $S$ lead to a closer agreement between $ARS$ of theory and experiment.
The need to use the experimental $S$ is due to the high share of aggregates (30.2\%) in this sample as pointed out above in the discussion on structural properties.
The occurence of aggregates is random, which shifts $S$ slightly towards unity and, for the case shown in Figure\,\ref{fgr:S_and_FF}d, increases scattering up to $\theta=70^\circ$.

We point out that many experimentally obtained disordered interfaces generally carry intrinsic correlations to some extend due to fabrication conditions and constraints, i.e. $S$, or any other appropriate representation of a structure in Fourier space, deviates from the value of a truly random structure.
As an example, it is known that black silicon surface textures prepared by dry-etching have certain correlation lengths that can be controlled through fabrication parameters, moreover, it has been shown that their correlation length correlates with the light-trapping abilities of these kind of textures in solar cells\cite{Steglich2014,Otto2015}.
However, as long as $S$ has no tendency towards zero for $q\rightarrow 0$ and $\bm{f}$ concentrates scattering into an angular region of small $\theta$'s, which is the case for most nanoparticles commonly used in the field of nanophotonics, these structures will still scatter mostly into the forward direction since their $\left|\bm{f}_0\right|^2$ and $S$ are both large. 
To suppress scattering into small $\theta$s, $S$ has to be as small as possible in the corresponding $q$-range, but large otherwise, which essentially describes $S$ of a nearly HuD structure. 
Therefore, we can identify (nearly) HuD as a key element to achieve high efficiency large-angle scattering.

Concerning array properties, specular transmission and reflection depend only on the areal nanodisk density $\rho$ but are independent of $S$, see Equations\,\ref{T} and \ref{R}.
This counter-intuitive result is a direct consequence of the first-Born approximation which we apply within the framework of our numerical approach, as it neglects any kind of coupling between scatterers that could change $\bm{f}$ in, e.g., dependence of distance to neighboring scatterers. 
However, throughout this work and also in Figure\,\ref{fgr:S_and_FF}e, where we plot $P_\text{sca}=1-R-T$ for different nearly HuD samples made of the same nanodisks but different $S$ and $\rho$, we find good agreement between numerical and experimental results. 
We take the observed agreement as a strong indication that the assumption of negligible inter-particle coupling in our samples is sufficiently justified. 
We attribute the deviations between the numerical and experimental values in case of the dense sample (orange) in Figure\,\ref{fgr:S_and_FF}e to the high number of aggregates in this sample.

However, since $P_\text{sca}$ depicts the amount of light scattered, but does not imply how the angular scattering response looks like, Figure\,\ref{fgr:S_and_FF}e also reveals the potential of (nearly) HuD structures to built highly efficient scattering interfaces with a scattering response on demand (see also \textit{supporting information}). 
The scattering response is up to the interplay between $\bm{f}$ and $S$ and can be tailored by adapting both precisely. 
It implies that one could change $\bm{f}$ by using another scatterer geometry, and $S$ by refining the fabrication process to suppress the occurrence of aggregates at high densities and enhance hyperuniformity, or use even another suitable process to enable the fabrication of desired $S$.

\section{Conclusion}
In summary, we achieved to prepare high-index nanodisk arrays of strong correlated disorder with a novel fabrication method.
The developed technique is scalable and allows control over important properties of the array, such as nanodisk height and diameter, and characteristic next-neighbor distance, through easy-to-access experimental parameters. 
Moreover, we could show that the disorder of the arrangement of the nanodisks is nearly hyperuniform, leading to rather unusual light scattering properties. 
While control over the angular scattering response of conventional scattering interfaces, either periodic or random, is essentially limited by either the structure or the form factor, near hyperuniform disorder enables both quantities to substantially impact the response. 
This particular ability, in consequence,  leads  to  a  fundamentally  novel  approach  to  tailor  light  scattering through tailoring both structure and form factor towards a scattering response on demand and paves the way to a new class of optical materials.

\section{Methods}

\subsection{Nanodisk array fabrication}
Standard 1\,mm thick 76\,x\,26\,mm object slides (soda-lime glass) were coated by thermal atomic layer (ALD) deposition in a Beneq TFS-200 ALD tool. 
First, the 13.8\,nm etch stop layer of Al$_2$O$_3$ was deposited by alternating pulses of trimethylaluminum (Al(CH$_3$)$_3$) and water (250\,ms pulse duration, 100 pulses per precursor, 3\,s purge) at 180$^\circ$C\cite{otto_alox2012}. Then, a 231.0\,nm layer of TiO$_2$ was deposited at 120$^\circ$C by 3633~alternating pulses of TiCl$_4$ and water (pulse duration 200\,ms).
This layer provides the material the nanodisks are made of.  
In more  detailed studies on the optical properties of ALD grown TiO$_2$ found in literature,\cite{aarik1995,aarik1997,saha2014} deposition temperatures below $\approx 165^\circ$C produce smooth, amorphous films with negligible optical losses in the visible spectrum compared to higher deposition temperatures, which agrees well with our observations by electron microscopy and spectroscopic ellipsometry. 
To complete the layer stack (please note Figure~\ref{fig:process}a), a layer of 19.5\,nm Al$_2$O$_3$ was deposited by ALD (165 cycles, otherwise same deposition parameters as above).

A dispersion of PMMA nanospheres with diameter of 499\,$\pm$\,10\,nm (microParticles GmbH, Berlin, Germany) was first treated in an ultrasonic bath for 30\,min to break up reversible aggregates and then diluted with a mixture of MicroPure$^{\text{TM}}$ water (18\,M$\Omega$cm) and 30\,\% isopropyl alcohol (p.a.\,grade). 
Potassium chloride (KCl) was used to carefully adjust the ionic strength of the dispersion (typically in the range of $10^{-5}$\,M to $10^{-3}$\,M).
The ALD coated substrates were exposed to the dispersion for 18 hours and rinsed in deionized water afterwards.
The samples were then rinsed in ethylene glycol multiple times to remove all water since Al$_2$O$_3$ surfaces can degrade in hot aqueous solutions during the subsequent tempering step \cite{tadanaga2008}.

Pattern transfer was performed via RIE in an Oxford Plasmalab 100 System.
For the Al$_2$O$_3$ etch process, BCl$_3$ and Ar were used in a ratio of 30:20 sccm at a pressure of 10 mTorr.
The inductively coupled plasma (ICP) power was 2500\,W and the capacitively coupled plasma (CCP) power was 63\,W, resulting in a dc-bias of -89\,V \cite{YunAlox2008}.
For the TiO$_2$ etch process the gas of CF$_4$:O$_2$:Ar was 40:6:10\,sccm with the pressure kept constant at 15\,mTorr. 
The ICP power was 1800\,W and the CCP power was 41\,W, resulting in a dc-bias of -81\,V.

 \subsection{Characterization}
For the microstructural characterization, SEM images were recorded with a field emission microscope (FEI Versa 3D) at 4 kV. 
Disk positions were extracted from images of 104\,x\,70\,$\mu$m$^2$ areas via template matching\cite{piechulla2018,opencv}.
Some disk pairs exhibit nominal surface-to-surface distances $\leq 0$\,nm (see e.g. Figure~\ref{fgr:Sq}g) due to slight merging of the PMMA particles during tempering.
Those pairs were defined as disks in aggregates.

The ALD layers were optically characterized on Silicon reference samples using a spectroscopic ellipsometer (J.A.~Woollaam M2000V). Due to the low absorption of TiO$_2$ the refractive index could reasonably approximated by Cauchy's equation with $n(\lambda_{\mu m})=2.281+0.024/\lambda_{\mu m}^2+0.006/\lambda_{\mu m}^4$, e.g. $n(\lambda=0.45\,\text{$\mu$m})\approx$\,2.54 and $n(\lambda=1.0\,\text{$\mu$m})\approx$\,2.31.\cite{aarik1997,saha2014}

Angular resolved scattering measurements were performed using a custom-made goniometer setup. 
White light of a halogen lamp was collimated, linearly polarized (where applicable) and directed onto the respective sample. 
The scattered light was collected and focused onto a fiber bundle by a single lens with an aperture corresponding to an acceptance angle of $\approx3.5^\circ$.
Accordingly, the lense-fiber setup was rotated about the sample from $\theta=4^\circ$ (to avoid recording the directly transmitted light) to $\theta=90^\circ$ in steps of $\Delta\theta=3^\circ$. 
A monochromator (Horiba iHR550) in conjunction with a silicon photodetector (Thorlabs DET210) and a lock-in amplifier (Stanford Research Systems SR830) was used to detect the light at the fiber output in the range of 450\,nm to 1000\,nm.
Directly transmitted light at $\theta=0^\circ$ without sample was recorded as reference $I_\text{ref}(\psi,\lambda)$ for each linear polarization angle $\psi$ and for unpolarized light, in order to characterize the system response.
The presented data are $I(\theta,\psi,\lambda)=(I_\mathrm{raw}(\theta,\psi,\lambda)-I_\mathrm{dark})/(I_\mathrm{ref}(\psi,\lambda)-I_\mathrm{dark})$.
Integrations over the forward-scattering half-space were $\int_{4^{\circ}}^{90^{\circ}} I(\theta,\psi,\lambda) \sin{\theta} d\theta $, where the $\sin{\theta}$ accounts for stronger representations of larger angles.  

\subsection{Numerical methods}
Point configurations were predicted via a random sequential adsorption scheme in a Monte Carlo type code.
The soft-sphere behavior of the charged colloidal particles was taken into account by modulating the probability of a particle to adhere on the substrate at a given random position within an extended radius of  neighboring particles\cite{piechulla2018}. 
Considering the extended radius, the saturation density in the experiment was slightly higher than in the hard-sphere case but approximately constant over a variety of number densities.
It could therefore be kept constant for the prediction of patterns.
Optimizations of the algorithm regarding computation time were inspired by the work of Zhang {\it et al.}\cite{zhangPreciseAlgorithmGenerate2013}

To calculate the angular resolved scattering response of a single nanodisk, the form factor $\bm{f(\hat{\bm{E}},\bm{k}_0,\bm{k}})$, we rigorously solved Maxwell’s equations in 3D with appropriate source, material, and boundary condition settings using the time-harmonic finite-element solver \textit{JCMsuite}\cite{Pomplun2007}.
To obtain the angular resolved scattering response of an infinitely extended array of identical nanodisks, $\bm{f}$ is combined with the structure factor $S(q)$ of the  point configuration of the particular array via our numerical approach based on Born's approximation (Equations\,\ref{T}-\ref{Born1}).  
Multipole decomposition was performed using an integrated function of \textit{JCMsuite} that applies the multipole expansion to the scattered fields and solves for the corresponding coefficients using an algorithm presented by Santiago \textit{et al.}\cite{Santiago2019}


\begin{acknowledgement}
This work was funded by the Deutsche Forschungsgemeinschaft (DFG) through program DFG-SPP 1839 “Tailored Disorder”, second period (RO 3640/6-2 and WE4051/19-2, project 278744673). 
The authors thank Claudia Stehr for her excellent technical support. 
\end{acknowledgement}

\bibliography{manuscript}
\newpage

\begin{suppinfo}

\subsubsection{Multipole expansion analysis}

\begin{figure}[h]
    \includegraphics[width=1.0\textwidth]{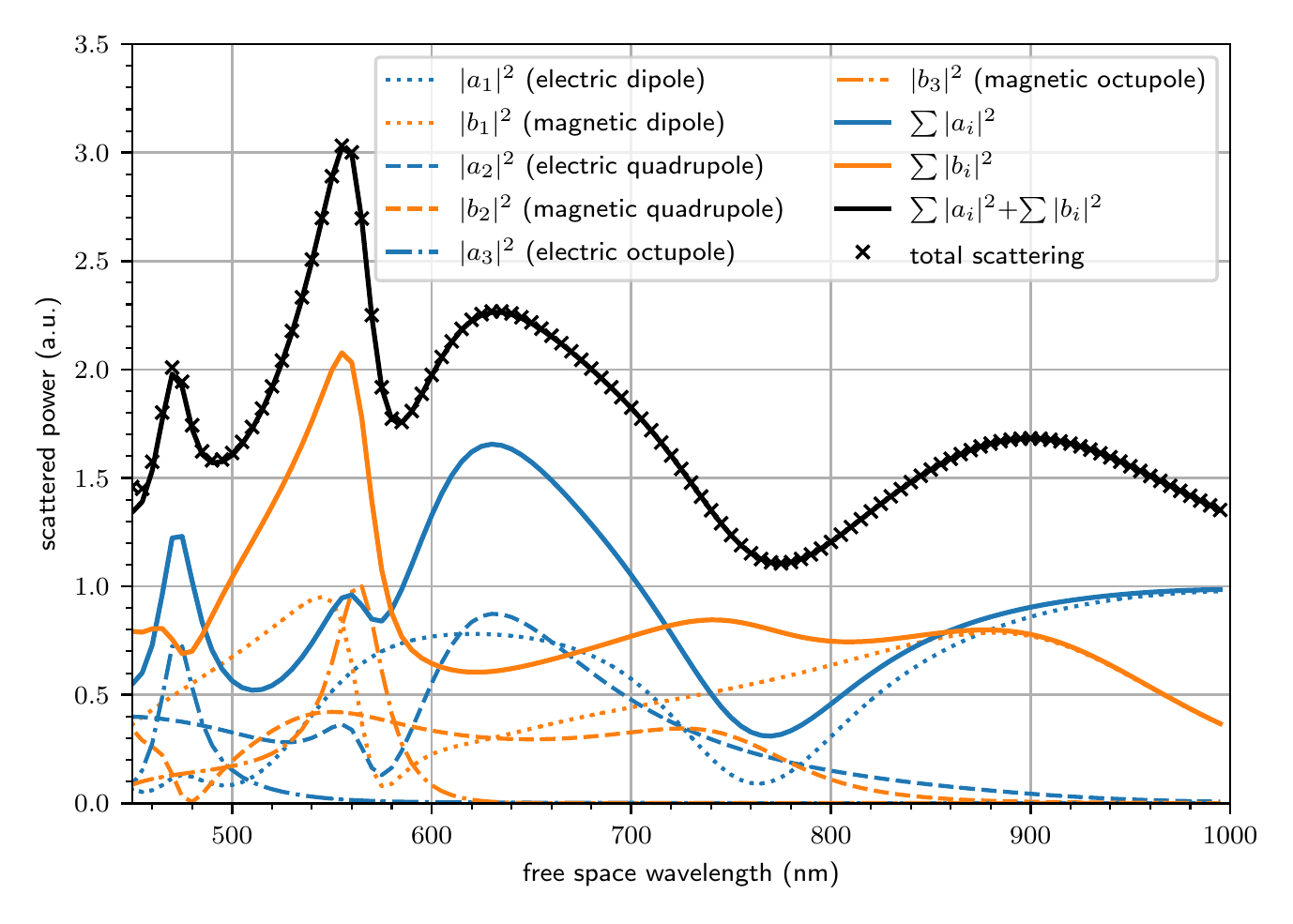}
    \caption{S1: Multipole expansion analysis of a single TiO$_2$ nanodisk in vacuum (diameter 455\,nm, height 231\,nm). 
    }
    \label{si:multipoles}
\end{figure}

In Figure\,\ref{si:multipoles}/S1 we plot the square modulus of the  expansion coefficients $a_i$ and $b_i$ of the electric and magnetic dipole ($i=1$), quadrupole ($i=2$), and octupole ($i=3$), respectively.
Multipoles of order higher than octupoles were found to be negligible in the spectral range considered. 
The sum of all contributions, $\sum (a_i + b_i)$, is in excellent agreement with the total scattered power derived by integration of the flux of scattered power integrated over a closed surface around the disk.

\subsubsection{Scattering efficiency}

\begin{figure}[h]
    \includegraphics[width=1\textwidth]{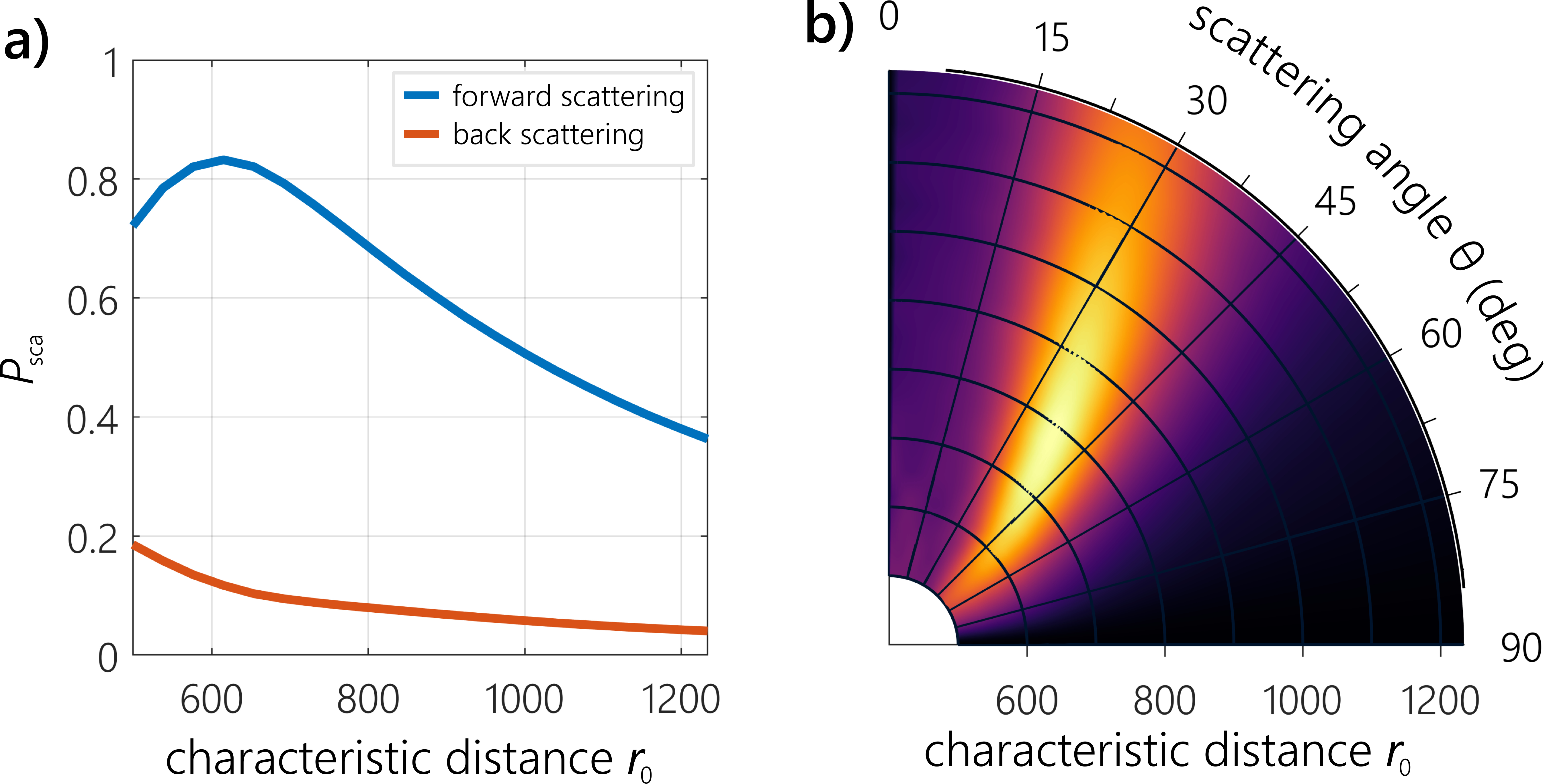}
    \    \caption{S1: Scattering at $\lambda=530$\,nm for nanodisk arrays with varying structure factors $S$ with characteristic distance $r_0$ that were numerically obtained with our modified RSA model. (a) Forward and back scattering and (b) corresponding $ARS$ (forward direction   ).
    }
    \label{si:efficiency}
\end{figure}

We produced a set of nearly Hud patterns with our modified RSA model and calculated the optical response using the corresponding  structure factors and the form factor that was used throughout this work. 
For a characteristic distance of $r_0=615$\,nm we find that at $\lambda=530$\,nm around 95\% of the irradiating light hitting the interface undergoes scattering, of which the major part, $P_\text{sca,f}=0.83$, is scattered into the forward direction, see Figure\,\ref{si:efficiency}a.
In Figure\,\ref{si:efficiency}b we plot the $ARS$ (forward direction) of these structures at the same wavelength.

\end{suppinfo}

\end{document}